\documentclass[12pt,tightenlines,nofootinbib,superscriptaddress,notitlepage, APS, pra]{revtex4-1}
\usepackage{amssymb}

\usepackage{amsfonts,amssymb,amstext,amsmath,amsthm,bm,slashed}
\usepackage[dvips]{graphicx}
\usepackage{mathtools}

\usepackage{fullpage}
\usepackage[title,titletoc]{appendix}

\usepackage{color}
\usepackage{float}
\usepackage{placeins}
\usepackage{bbold}
\usepackage{tabu}
\usepackage{verbatim}
\usepackage{comment}
\usepackage{enumerate}

\theoremstyle{definition}

\theoremstyle{plain}

\usepackage{relsize}

\newcommand{\be}{\begin{equation}}
\newcommand{\ee}{\end{equation}}
\newcommand{\barray}{\begin{array}}
\newcommand{\earray}{\end{array}}
\newcommand{\bea}{\begin{eqnarray}}
\newcommand{\eea}{\end{eqnarray}}
\newcommand{\bs}{\begin{subequations}}
\newcommand{\es}{\end{subequations}}
\newcommand{\beal}{\begin{align}}
\newcommand{\eeal}{\end{align}}

\newcommand{\bra}[1]{[ {#1}|}
\newcommand{\ket}[1]{|{#1}\rangle}
\newcommand{\mean}[1]{[{#1}\rangle}

\newcommand{\sket}[1]{| #1]}
\newcommand{\sbra}[1]{\langle {#1}|}
\newcommand{\smean}[1]{\langle {#1}]}
\def\rock{[ \xi |p|\xi]}
\def\rockX{[ \rho|\dot{x}|\rho]}

\def\sig{\sigma}

\def\eps{\epsilon}

\def\lam{\lambda}

\def\a{\alpha}
\def\eps{\epsilon}

\def\da{{\dot{\alpha}}}
\def\db{{\dot{\beta}}}
\def\dg{{\dot{\gamma}}}

\def\pa{\phantom{\alpha}}
\def\bs{\bar{\sigma}}

\def\bp{\bar{\pi}}

\def\br{\bar{\rho}}

\def\bxx{\bar{x}}
\def\bp{\bar{p}}
\def \bD{\bar{\Delta}}

\def \dr{\dot{\rho}}
\def\dxi{\dot{\xi}}

\newcommand{\bsy}[1]{\boldsymbol{#1}}

\newcommand{\rd}{\mathrm{d}}

\def\sl2c{SL(2,\mathbb{C})}

\newcommand{\mf}[1]{\mathfrak{#1}}

\newcommand{\abs}[1]{\vert #1 \vert}

\newcommand{\pb}[1]{\left\lbrace #1 \right\rbrace}
\newcommand*\overbar[1]{%
  \hbox{%
    \vbox{%
      \hrule height 0.5pt 
      \kern0.4ex
      \hbox{%
        \kern 0em
        \ensuremath{#1}%
        \kern 0em
      }%
    }%
  }%
}

\def \Tr {\mathrm{Tr}}

\def\bs{\bar{s}}

\def\bx {\bar{\xi}}

\def\bJ{\bar{J}}

\newcommand{\R}{\mathbb{R}}

\usepackage{bbm}
\def\Id{{\mathbbm 1}}

\def\R{{\mathbbm R}}

\newcommand{\mr}[1]{\mathrm{#1}}
\def\b{\beta}\def\b{\beta}

\allowdisplaybreaks

\usepackage{tikz}
\usetikzlibrary{decorations.pathmorphing}
\usepackage{tkz-euclide}
\usetikzlibrary{decorations.markings}
\tikzset{->-/.style={decoration={
  markings,
  mark=at position .5 with {\arrow{>}}},postaction={decorate}}}

\usepackage{cleveref}

\crefformat{equation}{#2eq.~(#1)#3}
\Crefformat{equation}{#2Eq.~(#1)#3}

\begin{document}
\title{A Classical and Spinorial Description of  the 
Relativistic Spinning Particle}

\author{{Trevor Rempel}}
\thanks{trempel@pitp.ca}
\affiliation{{Perimeter Institute for Theoretical Physics, Waterloo, Ontario, Canada, N2L 3G1}}
\affiliation{\small\textit{Department of Physics, University of Waterloo, Waterloo, Ontario, Canada, N2L 3G1}}
\smallskip
\author{{Laurent Freidel}}
\thanks{lfreidel@pitp.ca}
\smallskip 
\affiliation{{Perimeter Institute for Theoretical Physics, Waterloo, Ontario, Canada, N2L 3G1}}
\date{ \today}
\bigskip
\begin{abstract}
In a previous work we showed that spin can be envisioned as living in a phase space that is dual to the standard phase space of position and momentum. In this work we demonstrate that the second class constraints inherent in this ``Dual Phase Space'' picture can be solved by introducing a spinorial parameterization of the spinning degrees of freedom.
This allows for a purely first class formulation that generalizes the usual relativistic  description of spinless particles and provides several insights into the nature of spin and its relationship with spacetime and locality. In particular, we find that the spin motion acts as a Lorentz contraction on the four-velocity and that, in addition to proper time, spinning particles posses a second gauge invariant observable which we call proper angle. Heuristically, this proper angle represents the amount of Zitterbewegung necessary for a spin transition to occur. Additionally, we show that the spin velocity satisfies a causality constraint, and even more stringently, that it is constant along classical trajectories. This leads to the notion of ``half-quantum'' states which violate the classical equations of motion, and yet do not experience an exponential suppression in the path integral. Finally we give a full analysis of the Poisson bracket structure of this new parametrization. 
\end{abstract}

\maketitle
\section{Introduction}
Spin is an elusive physical entity. On one hand it can be understood as a purely quantum phenomena which represents internal degrees of freedom and provides a labeling of fields which is independent from their dynamics. On the other hand it is fundamentally intertwined with the concept of spacetime, being defined via irreducible representations of the Poincar\'e group. It is the relationship between spin and the concept of spacetime and localization which is at the heart of our interest in the subject. The working hypothesis being that insights into the nature of spin may give some clues about the quantum nature of spacetime, provided that we have fully understood the former. Achieving this aim requires a deep and refined handle on spin itself which, we propose, can be realized through an analysis of its classical representation. 

That one can describe spin classically stems from the fact, epitomized by Feynman, that any quantum phenomena can be described in terms of the phase $e^{iS/\hbar}$ gathered by the wave function $\psi$ when undergoing continuous motion. In the case of a massive relativistic particle the phase is well known to be given by $S=m\tau$; mass times the proper-time elapsed. The same is true for a relativistic  spinning particle, but  there exists an extra phase contribution due to the spin: $S= m\tau + 2\hbar s\phi$ where $\phi $ is the proper (spin) angle. Although the phase associated with motion through spacetime is clearly understood as a measure of proper time the same can not be said about the phase resulting from spin motion. This is what we seek to clarify in the present work. First, however, we need to understand the true spin degrees of freedom and their motion, only then will we be in a position to formulate a rigorous definition of $\phi$. As we will see, this story is made more complex, and interesting, by the fact that the spin motion affects spacetime motion in a non-trivial manner. In particular we find that the proper angle $\phi$ is a measure of the oscillation along the classical trajectory of a spinning particle, a phenomena known as Zitterbewegung \cite{schrodinger_1930}. We also show that the measure of proper-time is affected by the spin motion, experiencing a Lorentz like contraction when the particle undergoes a spin transition. The precise disentangling of Zitterbewegung from a spin transition is one of the main results of our paper.

Classical, realistic models of spin in which the spin degrees of freedom are real valued and commuting\footnote{As opposed to  models which utilize Grassmann variables.} have a long and vibrant history, extensively outlined in \cite{rempel_2015} and references therein (see also \cite{Frydryszak:1996mu}). Of these it is sufficient to focus on those which use group theoretic methods in their derivation of a classical action. The first to present such a model  were Hanson and Regge \cite{hanson_1974} who assumed that for each value of some evolution parameter $\tau$, the spinning particle is characterized by an element of the Poincar\'e group $(x^\mu(\tau), \Lambda^\mu_{\pa \nu}(\tau))$.  One can then write the Lagrangian in terms of  Poincar\'e invariant quantities after imposing a set of orthogonality or Dixon conditions. A much simpler model was proposed by Balachandran \cite{balachandran_1976, balachandran_1979, balachandran_1981} who also assumed that the configuration space of a spinning particle should be identified with the Poincar\'e group. Although these initial models were useful they were somewhat ad-hoc; a formal application of group theory to classical systems required the advent of a powerful mathematical framework known as the coadjoint orbit method.

Proposed by Kirillov \cite{kirillov_1976}, Kostant \cite{kostant_1970} and Souriau \cite{souriau_1970,souriau_1997}, the coadjoint orbit method is based on a theorem showing that the coadjoint orbits of a group form a symplectic manifold and therefore have a natural interpretation as the classical phase space of some system. A number of authors have subsequently made use of this method to propose classical descriptions of the spinning particle, principally by considering its application to the Poincar\'e group. A particularly concise model due to Wiegmann \cite{wiegmann_1989B} uses a single four-vector $n_\mu$, identified as an element of $S^2$ in a particular frame, to encode the spinning degrees of freedom.  The main idea behind these models, explicitly spelled out by Wiegmann, is to encode the spin degrees of freedom into the analog of a Wess-Zumino term carrying the spin information.

Unfortunately, the models mentioned above suffer from a serious limitation, namely that they are only valid on solutions of the mass-shell constraint, i.e. they describe physically asymptotic states. As such, they are not suited to the worldline description of loop amplitudes and therefore can not accommodate fully quantum processes. Attempts to overcome this restriction gave rise to a convoluted series of models which sought to extend the description of spin to the full and unconstrained particle phase space. The primary hurdle was to express the orthogonality constraints that entangle spin and spacetime in a physically meaningful and mathematically tractable form. 
There have been two trends apparent in these attempts, one where the spin degrees of freedom are represented by vectors and the other where they are represented by spinors.  In vectorial models \cite{wiegmann_1989B,Zakrzewski:1994bd,Deriglazov:2012kw,Plyushchay:1990cz} one obtains a set of second class constraints whose form depends on the vectors are chosen and these do not necessarily possess a clear physical or geometrical meaning. On the other hand, the spinorial formulation allows one to reduce the number of the constraints, even to the point of eliminating second class constraints entirely. This approach was initiated by Penrose \cite{Penrose:1987uia} and subsequently developed by Bengtsson \cite{Bengtsson:1987ap} for massless spinning particles. It was then extended  to the massive case by several authors \cite{lyakhovich_1994,Kassandrov:2009jd,barut_1984B,pavsic_1992}. A particularly notable spinor model, and one that will be important for us, is that of Lyakovich et al. \cite{lyakhovich_1994}, further developed in \cite{lyakhovich_1996}. It is claimed to be a ``universal description'' of the spinning particle including both massive and massless particles, as well as those with continuous spin. This model has also been generalized to  spinning particles in any dimension, see \cite{lyakhovich_1998A} and \cite{lyakhovich_1998B}.

One of the goals of this paper, together with \cite{rempel_2015,rempel_2016}, is to linearize the history discussed above and present a self contained description of the relativistic spinning particle from first principles. In the first paper \cite{rempel_2015} we started from the coadjoint orbit formalism and showed that the relativistic spinning particle can naturally be recast in a vectorial representation called the ``Dual Phase Space'' (DPS) model in which the spin degrees of freedom carry their own phase space. This allowed us to introduced, for the first time, the interaction vertex in terms of 
a realistic worldline interpretation as a dual locality condition. In the second work \cite{rempel_2016} we showed that the DPS picture  
implies that relativistic spin can be understood physically as an entangled pair of constituent particles and could be quantized as such.
In this work we address the issue of resolving fully all the second class constraints present in the original DPS model. We show that there is a unique resolution that agrees with Lyakovich et al.  and we  give a full description of how spin motion affects spacetime motion. 
\section{Overview}
The "Dual Phase Space" model parameterizes the phase space of the relativistic particle of mass $m$ and spin $s$ in terms of two pairs of canonically conjugate four vectors $(x^\mu, p^\mu)$ and $(\chi^\mu, \pi^\mu)$. The pair $(x^\mu, p^\mu)$ represent the standard position and momentum of the particle while $(\chi^\mu, \pi^\mu)$ are ``dual'' variables which encode the spinning degrees of freedom. The dynamics of the particle are then determined by a set of six real constraints, and if the particle is massive four of these are second class. These constraints can be presented most straightforwardly by introducing a complex ``spin'' vector
\begin{equation}
\ell^\mu=  \frac{\pi^\mu}{\eps} + i s\frac{\chi^\mu}{\lam},
\end{equation}
where $\lambda$ and $\epsilon$ are fundamental length and energy scales respectively and satisfy $\lambda \epsilon =\hbar$.

The first class constraints which define mass and spin are simply restrictions on the length of the momenta and spin vector
\be
p^2+m^2=0, \qquad \ell \ell^* = 2 s^2.
\ee
These are then supplemented by additional constraints which form a second class system.
The first result of this paper is to construct a purely first class model by using the spinor formalism to solve the second class constraints. The purpose of such a re-parameterization is two fold, first it provides, as already mentioned,  a greater control over the action while allowing for a better understanding of the effect of spin on particle dynamics. Secondly, it makes a connection with the standard Dirac formalism and therefore should permit a description of fermions\footnote{It was shown in \cite{rempel_2016}  that, upon quantization, DPS only yields integer spins.}.
  
We find that the general solution to the second class constraints is obtained by setting
\begin{align}
\ell  = \frac{ \ket{\xi}\bra{\xi} p }{m},
\end{align}
where $\xi_\alpha= \ket{\xi}$ is a spinor, $\bra\xi =\epsilon^{\alpha\beta} \xi_\beta$ is the transposed spinor,
and $p$ is the momenta represented as a $2\times 2$ hermitian operator. The resulting first order action has two undetermined Lagrange multipliers corresponding to the mass shell and spin constraint. 
We can interpret the Lagrange multipliers as generators of two gauge invariant quantities, proper time $\tau(t)$ (dual to the mass shell) and proper angle $\phi(t)$ (dual to the spin constraint) and we show that the action is of the form 
\be \label{S1}
S= m\tau(t)+ 2\hbar s \phi(t).
\ee

The spin velocity $\ket{\dot\xi}$ can be expressed as a function of two complex coefficients $(\mf{a},\mf{b})$ that characterize the spin motion and which are defined by the expansion
\be
\ket{\dot{\xi}}= 
\mf{a} \ket{\xi} 
+\frac{\mf{b}  m }{2\hbar s}\,
{\dot{x}} \sket{\xi}.\qquad
\ee
The proper time and the proper angle are then explicitly given by 
\be\label{S2}
\dot{\tau} = |\dot{x}| \sqrt{1-\mf{b}^2}, \qquad \dot{\phi}=  {\mathrm Im}(\mf{a}).
\ee
where $|\dot{x}|\equiv \sqrt{-\dot{x}^2}$. 
The fact that the action is independent of ${\mathrm{Re}(\mf{a})}$ means that it is invariant under spin rescaling $\ket\xi \to \alpha \ket\xi$, with $\alpha\in \mathbb{R}^+$ which is  
essentially the expression of Lorentz invariance from the spin point of view.
 Comparing this to the standard action for the spinless relativistic particle one notices that the four-velocity is modified by a factor of the form
$\sqrt{(1 - \mf{b}^2)}$.  This shows that spin motion can be viewed as inducing a Lorentz contraction of the four-velocity! In addition, one notes that there is a maximal speed of spin propagation encoded into the causality condition $|\mf{b}| \leq 1$. Violating this bound would yield a spacelike velocity $\dot{x}^2 > 0$. Note that the parameter $\mf{b}$ measures the propensity of spin to flip along the motion of the particle. 

 Further analysis reveals an even more stringent restriction on the  classical spin motion: If a relativistic spinning particle has an initial configuration given by $(x, \xi)$ and $x'$ is in the future light cone of $x$ , then there is a classical path connecting $(x,\xi)$ and $(x',\xi')$ only if $\xi' = \xi$, that is only if the spin state does not evolve under classical motion. It follows that there are trajectories which have $\dot{\xi} \neq 0$ but which still satisfy the causality constraint $|\mf{b}|\leq 1$. These ``half-quantum'' states are interesting because although they are not classical they are not exponentially suppressed in the path integral either.  This possibility explains why spin and its motion can only be fully understood as a quantum object since the boundary between quantum and classical is not as sharply defined as it is for spacetime motion.

We complete the first order formulation by computing explicitly the commutators among the position and spin variable and we witness that the presence of spin renders the position variable non-commutative (as already noticed in \cite{Zakrzewski:1994bd,Freidel:2007qk,Das:2009se}). The calculation is involved but is simplified by considering the symmetries of the symplectic potential/form. In particular we find that the position coordinates acts a type of boost generator on the spin variables:
\be 
\pb{\xi_\a, x_{\b \db}} = \frac{p_{\a \db}\xi_\b }{m^2}.
\ee
From our analysis we can clearly see two new phenomena associated with spin, the existence of a spin causality constraint and the possibility of``half-quantum'' states.

\section{First Class Formalism}
In the DPS model the phase space of the relativistic spinning particle is parameterized by two pairs of canonically conjugate four vectors $(x_\mu, p_\nu)$ and $(\chi_\mu, \pi_\nu)$ with Poisson brackets
\be
\pb{x_\mu, p_\nu}=\hbar \eta_{\mu\nu},\qquad
\pb{\chi_\mu,\chi_\nu}=\hbar \eta_{\mu\nu}. 
\ee
The fundamental length and energy scales $\lambda$ and $\epsilon$ 
satisfy  $\lambda \epsilon =\hbar$, they allow for the unification of the ``dual'' position $\chi_\mu$ and ``dual'' momenta $\pi_\mu$ into a single complex vector
\be
\ell_\mu \equiv  \frac{\pi_\mu}{\eps} + i s\frac{\chi_\mu}{\lam},
\qquad \pb{\ell_\mu,\ell^*_\nu}= 2s \eta_{\mu\nu}.
\ee
The dynamics of the spinning particle are then characterized by two sets of constraints with the first being obtained from a simple restriction on the lengths of $p$ and $\ell$:
\be
p^2 + m^2 = 0, \quad \ell \ell^* = 2s^2. \label{sp:first_class}
\ee
The remaining constraints are then given by two pairs of orthogonality conditions 
\be
p\cdot \ell = 0,
\qquad \ell^2 = 0,
\label{sp:second_class}
\ee
and their conjugates.  The relationship between this formulation and the  coadjoint orbit formalism is detailed in \cite{rempel_2015}. The main point is that the total angular momenta can be expressed as a sum of the orbital angular momenta $L_{\mu\nu} =  (p\wedge x)_{\mu\nu}$ and the spin angular momenta $S_{\mu\nu}= (\pi \wedge \chi)_{\mu\nu}$.

It is easy to verify that the constraints in \cref{sp:first_class} are first class. However, although the constraints in \cref{sp:second_class} commute with each other they do not commute with their conjugates unless $m$ or $s$ vanish. The first major result of this paper is to solve this second class system by means of the spinor formalism leaving us with a purely first class representation of the relativistic spinning particle. For simplicity we restrict our analysis to four dimensional spacetime since this allows us to utilize two-component spinors.

Let us begin with the null condition $\ell^2=0$, which is solved by noting that any complex null vector can be represented by a product of spinors $\xi_\a$, $\bar\zeta_\da$, viz
\begin{align}\label{nullsolve}
\ell_{\a\da} =  \xi_\a \bar{\zeta}_\da,\qquad  \ell =\ket{\xi}\sbra{\zeta}.
\end{align}
Here and in what follows we will utilize the spinor formalism quite extensively (see \cite{Penrose:1987uia,Dreiner:2008tw, Martin:1997ns, britto1}). For the readers' convenience we include a brief overview here, for a more detailed presentation see Appendix \ref{sf:app:spinor}.

Denote by  $\chi^\a$, $\alpha=0,1$, a two-dimensional complex spinor and $\bar{\chi}^{\da} = (\chi^\a)^\dagger$ its complex conjugate. Indices are raised and lowered with the epsilon tensor $\epsilon^{\a\b}$ which is the skew symmetric tensor normalized by $\epsilon^{01}=1$, i.e.
 \begin{align*}
\chi^\a = \eps^{\a\b}\chi_\beta, \quad \chi_\a = \eps_{\a\b}\chi^\b, \quad \bar{\chi}^\da = \eps^{\da\db}\bar{\chi}_{\db}, \quad \bar{\chi}_{\da} = \eps_{\da\db}\bar{\chi}^\db.
\end{align*}
These quantities are represented as bras and kets (see also \cite{Freidel:2013fia}) via
\begin{align}
\ket{\chi} &= \chi_\a, & \bra{\chi} &= \chi^\a, &
\sket{\chi} &= \bar{\chi}^\da, &  \sbra{\chi} &= \bar{\chi}_\da,
\end{align}
with the notation being specifically designed to distinguish a spinor from its conjugate. Note also that we have adopted a convention in which the epsilon tensor satisfies $\eps_{\a\gamma}\eps^{\gamma \beta} = \delta^\b_\a$.  The SL$(2,\mathbb{C})$ invariant contractions between spinors are denoted by a rocket:
\begin{align}
\mean{\zeta | \xi} :=   \zeta^\a \xi_\a, \qquad \smean{\zeta |\xi} := \bar{\zeta}_\da\bar{\xi}^\da,\qquad   \smean{\zeta |\xi}= -\mean{\zeta | \xi}^*.
\end{align}
Let $(\sig ^a)_{\a\da} = (\Id_{\a\da}, \vec{\sig }_{\a\da})$ be the standard four vector of sigma matrices, and  $(\bar{\sig }^a)^{\da\a} \equiv (\sig ^a)_{\b\db}\eps^{\a\b}\eps^{\da\db}$ the same vector but with indices raised. Given a real vector $p^a$ we can construct  two by two hermitian  operators
$p_{\a\da}: = p_a (\sig ^a)_{\a\da} $ and $ \bar{p} = p_a \bar{\sig }^a$ as well as the hermitian pairing 
$\bra \xi p \sket{\zeta}= \sbra{\zeta}\bar{p}\ket{\xi}$. This completes the brief introduction to the spinor formalism.
 
Given the parameterization in \cref{nullsolve} the remaining second class constraint is equivalent to $\sbra{\zeta}\bar{p}\ket{\xi} = 0$ which has the general solution $\sbra{\zeta}\bar{p} \propto \bra{\xi}$. The normalization can be chosen arbitrarily and to keep the spinor dimensionless we put $m \sbra{\zeta} =   \bra{\xi} p$, provided that $m \neq 0$. Thus, the general solution of \cref{sp:second_class} is 
\begin{align}
\ell_{\a\da}=\frac{ \xi_\a\xi^{\b} p_{ \b \da}  }{m}\qquad\mathrm{or}\qquad \ell  = \frac{ \ket{\xi}\bra{\xi} p }{m}.
\end{align}

The first class constraints, \cref{sp:first_class}, are expressed in terms of these new variables as 
\be 
\Phi_m = \frac12\Tr(p\bar{p}) -m^2,\qquad
\Phi_s = \frac12 \rock -ms.
\ee 
To obtain an action we simply add these constraints to the symplectic potential $\Theta = p_\mu dx^\mu + \pi_\mu d\chi^\mu$. As a one-form on phase space the symplectic potential can be evaluated on the second class constraints \cref{sp:second_class} and the spin part $\pi_\mu d\chi^\mu$ expressed entirely in terms of the spinor variable $\xi^\alpha$, viz
\begin{align}\label{sf:symplectic_concise}
\Theta = -\frac{1}{2}\Tr(\bar{p} d{x}) +\frac{i\hbar}{2m} \frac{\bra{\xi}{p}\sket{ \xi} }{ 2 m s } \,\bar{p}^{\a \da} \left(
    \xi_{\a} \rd\bx_{\da}-\bx_{\da} \rd \xi_{\a} \right).
\end{align}
It is interesting to see that the symplectic structure already depends on a unit of  action through $\hbar$ even if it is a classical entity. This expresses mathematically the notion that spin blurs the sharp distinction between classical and quantum that we are familiar with. Without loss of generality and up to a redefinition of Lagrange multipliers we can implement the first class constraints in $\Theta$. It will therefore be convenient to work with the simpler version
\be
\label{sf:symplectic2}
\Theta_{m,s} = -\frac{1}{2}\Tr(\bar{p} d{x}) +\frac{i\hbar}{2m} \,\bar{p}^{\a \da} \left(
    \xi_{\a} \rd\bx_{\da}-\bx_{\da} \rd \xi_{\a} \right),
\ee
which summarizes the symplectic structure of the relativistic spinning particle.

\section{Classical Action for the Relativistic Spinning Particle}

\subsection{First Order Action and Half-Quantum States}

The action $S_{m,s} = \int_0^\tau dt L_{m,s}(t)$ for the relativistic spinning particle has Lagrangian $L_{m,s} = \Theta_{m,s} + \frac{N}{m}\Phi_m + \frac{M}{m}\Phi_s$. 
Introducing the quantity $\theta(\xi): = i\hbar\left( \ket{\dot\xi}\sbra{\xi}-\ket{\xi}\sbra{\dot\xi}\right)$ we find that the Lagrangian is expressed explicitly as
\be\label{sf:lagrangian}
L_{m,s}= -\frac1{2 m} \Tr\left( \bar{p} \left( m\dot{{x}} +{{\theta}}(\xi)- M \ket{\xi} \sbra{\xi}\right)
-\frac{N}2 p\bar{p} 
\right)  -\frac{Nm}{2} - {M s}.
\ee
The first class constraints $\Phi_m$ and $\Phi_s$ generate time translations (parameterized by $\a$) and local spin rotations (parameterized by $\b$) respectively. Both of these gauge transformations leave the Lagrangian invariant and act on the phase space variables as
\be\label{sf:time_trans}
\delta_{(\alpha,\beta)} N := \dot\alpha,\qquad
\delta_{(\alpha,\beta)} M := \dot{\beta},\qquad  \delta_{(\alpha,\beta)} x := \frac{\a p}{m},\qquad
\delta_{(\alpha,\beta)} \ket{\xi}:= -\frac{i \b}{2\hbar}\ket{\xi},
\ee
while $\delta_{(\alpha,\beta)}p=0$ and $\delta_{(\alpha,\beta)} \theta(\xi) = \dot\beta \ket{\xi} \sbra{\xi}$.
These transformations can also be used to fix the Lagrange multipliers $N$ and $M$ to constant values, giving rise to two gauge invariant observables, the {\it proper time} $\tau$ and the {\it proper angle} $\phi$:
\be
\tau(t) =  \int_0^t \rd t' N(t'),\qquad
\phi(t) = \frac1{ 2 \hbar} \int_0^t \rd t' M(t'). 
\ee
The appearance of a new type of observable in addition to proper time is one of the most relevant facts about spin from the perspective developed here.

To obtain the first order action we need to solve the equation of motion for $p$ which is given by $N p= m\dot{x} + \theta(e^{i\phi}\xi)$. Inserting this into the Lagrangian we find
\be\label{firstorder}
L_{m,s} = -\frac1{4N m} \Tr \left[\left(m\dot{{x}} + {{\theta}}(e^{i\phi}\xi)\right)\left(m\dot{\bar{{x}}} + \bar{{\theta}}(e^{i\phi}\xi)
\right)\right]  -\frac{Nm}{2} - 2{\hbar s} \dot \phi,
\ee
where we have used that $2 \hbar \dot{\phi}= M$. We can further expand $L_{m,s}$ by means of the identities
 \be\label{sf:importantminus}
\Tr\left[\dot x \bar\theta(\xi)\right]
= -2 \hbar {\mathrm{Im}}\left( \bra{\dot \xi} \dot x\sket{\xi} \right)\qquad\mr{and}\qquad
\Tr\left[\theta(\xi)\bar\theta(\xi)\right] = - 2\hbar^2 |\mean{\xi|\dot\xi }|^2,
\ee
while also making use of the re-parametrization invariant spinor velocity 
 $ \ket{\partial_\tau\xi} = \dot{\ket\xi} / N$. We find
\be\label{sf:action}
L_{m,s} =
\underbrace{\frac{1}{2\tilde{N}}  \dot{x}^2 -
\frac{\tilde{N}}{2 }\left(m^2 - |\hbar \mean{\xi|\partial_\tau\xi }|^2\right)}_{\mr{Modified \; Mass-shell}}
+\overbrace{{  \hbar } \, {\mathrm{Im}}\left(\bra{\partial_\tau \xi} \dot x\sket{\xi} \right)}^{\mr{Spin\; Potential}}
+ \underbrace{{\hbar}\dot{\phi}\left( \bra{ \xi} \dot{x}\sket{\xi} - 2s\right)}_{\mr{Spin\; Constraint}},
\ee
where we have defined $\tilde{N}\equiv N/m$. Written in this form, the Lagrangian is valid in the massless limit as well. 

As seen above there are three terms which make up the Lagrangian: A modified mass-shell with effective mass $M$ given by 
\be\label{mass}
M^2 = m^2 -|\hbar \mean{\xi|\partial_\tau\xi }|^2, 
\ee a potential that couples the linear velocity to the spin, and finally the spin constraint 
\be\label{spinc}
\bra{ \xi} \dot{x}\sket{\xi}=2\hbar s.
\ee 
The minus sign appearing in the modified mass-shell, \cref{mass}, imposes a {\it causality constraint}: At a classical level the linear velocity must be timelike or null, i.e $M^2 \geq 0$, and so the spin motion must satisfy
\be\label{sf:spinCausality}
\hbar |\mean{\xi|\partial_\tau \xi }| \leq m.
\ee
Therefore, while the component of $\dot{x}$ along $\sket\xi$ is fixed by \cref{spinc}, the causality constraint restricts the spin  velocity $|\mean{\xi|\dot\xi }|$ to be bounded from above. Of course, this is a classical restriction and can be violated at the quantum level. These are the virtual processes whose amplitudes will be suppressed in the path integral.

We can extend this analysis to the semi-classical level and see more clearly the delineation between which processes will experience an exponential suppression and those which will not. Specifically, let us examine the trajectories defined by the classical equations of motion. For $x$ we find that the evolution is characterized by
\be \label{xevolv}
m\dot{x} + \theta ( \xi)  = N P+ \dot{\phi} \ket\xi\sbra\xi,
\ee 
where $P^\mu$ is a constant of motion. It follows that the particle will in general undergo oscillatory motion, known as Zitterbewegung \cite{schrodinger_1930}, due to the rotation of the spin, on top of the standard linear evolution. On the other hand, the equation of motion for $\xi$ reduces to\footnote{See Appendix \ref{sf:app:first} for more details.} $\xi(\tau) = e^{i\phi(\tau)}\xi(0)$ which yields an even more stringent restriction on the spin motion than \cref{sf:spinCausality}. In particular, it implies that the spin state can only change by an internal phase during classical evolution, hence $\theta ( \xi)= \dot{\phi} \ket\xi\sbra\xi$ and so Zitterbewegung motion is not experienced. 

 Observe that it is possible to violate the restriction on the spin evolution while still satisfying both the causality constraint \cref{sf:spinCausality} and the classical equation of motion \cref{xevolv} for $x$. This means that this is possible to have a quantum trajectory which permits spin flips, i.e.
 \be
 \mean{\xi(\tau)| \xi(0)} \neq 0.
 \ee 
Such ``half-quantum'' states represent trajectories which are not fully classical yet will not be exponentially suppressed in the path integral. 
More precisely, these trajectories will necessarily experience Zitterbewegung since by definition $\theta(\xi)\neq \dot{\phi} \ket\xi\sbra\xi$. However, even though this type of spin motion is not classical it is still possible to have a solution of \cref{xevolv} which satisfies the spin causality constraint. Hence we do not expect to see an exponential supression for these transitions. 

 Normally motion is classified as either classical, in which case the classical equations of motion are satisfied, or quantum, in which case the classical action is imaginary. Little is known about ``half-quantum'' states and they deserve further exploration; it is possible that they represent some form of  entanglement.\\

\subsection{Second-Order Action}

The second order action can be obtained from \cref{sf:action} by integrating out $\tilde{N}$ and $\dot{\phi}$. For $\tilde{N}$ we proceed in the usual fashion by solving its equation of motion,
\be
\tilde{N}^2 = -\frac{\dot{x}^2}{ \left(m^2 -
|\hbar \mean{\xi|\partial_\tau \xi }|^2\right)}
\ee and substituting the result back into $L_{m,s}$. 
The integration over $\phi$ on the hand imposes the spin constraint \cref{spinc}. 
In order to solve it we introduce a spinor $\rho_\a$, free of constraints, and which is related to $\xi$ via\footnote{We assume that $\bra{\rho}\dot x \sket{ \rho}>0$.} 
\be
\sket\xi 
= \sket \rho \, \sqrt \frac{{2 N  s} }{ {\bra{ \rho} \dot x\sket{\rho}} } .
\ee
Combining these transformations gives
the second order action
\be\label{sf:Actionf}
S = m\int_0^1\rd\tau \sqrt{-\dot x^2 \left( 1 -
\left|\frac{2\hbar s }{m}\frac{\mean{\rho|\dot\rho }}{\bra{\rho}\dot x \sket{ \rho}}\right|^2 \right)}
+ 2\hbar s 
 \int_0^1\rd\tau \left( \frac{{\mathrm{Im}}\bra{\dot \rho} \dot x\sket{\rho} }{\bra{\rho}\dot x \sket{ \rho}} \right).
\ee
As in the first order case, this action is invariant under time translations and spin rotations, now expressed as 
\be
\delta_{(\alpha,\b)} x =\a \dot{x},\qquad \delta_{(\alpha,\b)} \ket{\rho} = \a \ket{\dot{\rho}} + i\beta  \ket{\rho},
\ee
where $\a,\b\in \R$. The proper time and proper angel can be identified with the first and second term of \cref{sf:Actionf} respectively, viz.
\bea
 \tau(t)&\equiv&\int_0^t \rd t' \sqrt{-\dot x^2 \left( 1 -
\frac{|2\hbar s\mean{\rho|\dot\rho }|^2}{ m^2\bra{\rho}\dot x \sket{ \rho}^2} \right)},\label{sf:properTime}\\
\phi(t) &\equiv&  {\mathrm{Im}} \int_0^t\rd t' \left( \frac{\bra{\dot \rho} \dot x\sket{\rho} }{\bra{\rho}\dot x \sket{ \rho}} \right).\label{sf:properAngle}
\eea
\subsection{Decomposing Spin Velocity}
The spinors $\ket{\rho} $ and ${\dot{x}} \sket{\rho}$ form a basis for spinor space provided that $\bra{\rho}\dot x \sket{ \rho}\neq 0$.  Therefore, we can expand the spin velocity in this basis by introducing two complex functions $(\mf{a}(\tau),\mf{b}(\tau))$, with $|\mf{b}|<1$:
\be
\ket{\dot{\rho}}= 
\mf{a} \ket{\rho} 
+\frac{\mf{b}  m }{2\hbar s}\,
{\dot{x}} \sket{\rho}.
\ee
It is straightforward to solve for $\mf{a}$ and $\mf{b}$
\begin{align}\label{mfaANDmfB}
\mf{a} = \frac{\bra{\dr}\dot{x}\sket{\rho}}{\rockX}, \qquad \mf{b}= 
\frac{2\hbar s}{m}\frac{\mean{\rho | \dr}}{ \rockX},
\end{align}
from which it follows that 
\begin{align}
S = m \int d\tau |\dot{x}| \sqrt{1-|\mf{b}|^2} + 2\hbar s \int d\tau \mr{Im}(\mf{a }).
\end{align}
We see that  knowledge of the spin velocity at all times uniquely determines the proper time and proper angle. In particular, if the spin velocity has a component along $\dot{x}\sket{\rho}$ 
the proper time runs at  a slower pace and so $\sqrt{1-\abs{\mf{b}}^2}$ can be viewed as a time contraction factor due to the spin motion.
In Appendix \ref{sf:app:second} we extend this analysis a bit further and derive the 
 equations of motion associated with the second order action \cref{sf:Actionf}. 
The action in \cref{sf:Actionf} is a special case of the one derived by Lyakhovich et. al. in \cite{lyakhovich_1996}. The difference between the two comes from the inclusion of a term in Lyakovich's model which allows for the description of continuous spin particles (CSP's). As DPS is equivalent to the restricted version of the latter model (as established in this paper) it is reasonable to assume that there is a generalization of the Dual Phase Space model which will also permit the inclusion of CSP's. We will explore this possibility more fully in a subsequent paper.

\section{Poisson Brackets}
Computing the Poisson algebra is rather tedious but can be simplified somewhat by first considering the symmetries of the symplectic potential/form. From \cref{sf:symplectic_concise} we have that the symplectic potential is expressed in-terms of the original spinor $\xi_\a$ as
\be\label{sf:symplectic}
\Theta = -\frac12 \Tr( p  \rd \bar{x}) +  \frac{i\hbar}{2m}\frac{\bra{\xi}p
\sket{\xi}}{2ms} 
\left(\bra{\xi}p\sket{\rd\xi} -\bra{\rd \xi}p\sket{\xi}\right).
\ee
The symmetry group of $\Theta$ is the Poincar\'e group, which factors as the semi-direct product of the translation group and the group of left and right rotations.
Let the infinitesimal generators of right and left rotations be denoted by $\rho_\a{}^\b$ and $\bar{\rho}^\da{}_\db$, respectively. Under these rotations the phase space variables transform in the following  manner:
\begin{align}
\delta_\rho^R {x} &=  \rho x,  & \delta_\rho^R p &=  \rho p, &
\delta_\rho^R \ket{\xi}&= \rho\ket{\xi},& \delta_\a^R\sbra{\xi}&= 0,\label{sf:rightlow}\\
\delta_{\br}^L {x} &= x\br,  \,  & \delta_{\br}^L p &= p\br, &
\delta_{\br}^L \ket{\xi}&=0,& \delta_{\br}^L\sbra{\xi}&=\sbra{\xi}\br \label{sf:leftlow}.
\end{align}
We do not require $\rho$ or $\br$ to be traceless and so rotations have a non-trivial action on the epsilon tensor, in particular
\begin{align}
\delta^R_\rho \eps_{\a\b} = \rho_\a{}^\gamma \eps_{\gamma \b} + \rho_\b{}^\gamma \eps_{\a\gamma}, \qquad \delta^R_\rho \eps^{\a\b} = -\eps^{\gamma \b}\rho_\gamma{}^\a -\eps^{\a\gamma} \rho_\gamma{}^\b, 
\end{align}
where the second equality follows by demanding invariance of $\delta^\a{}_\b$. Identical results hold for left rotations of $\bar{\eps}$. Thus, the action of left and right rotations on quantities with raised indices can be obtained from \crefrange{sf:rightlow}{sf:leftlow} by adding a minus sign and moving the rotation matrix to the other side, e.g. $\delta_\rho^R \bra{\xi} = -\bra{\xi} \rho$, which implies that the rocket $\mean{\xi|\xi}$ is SL$(2,\mathbb{C})$ invariant.  We can also denote the infinitesimal generator of translations as $a_{\a\da}$, which acts only on the positional coordinate $x$ as $\delta_a x = a$. The Hamiltonian vector fields associated with these transformations are given by
\begin{align}
R_\rho &\equiv -(\bar{x}\rho)^{\da\a}\frac{\partial}{\partial \bar{x}^{\da\a}} + (\rho p)_{\a\da}\frac{\partial}{\partial p_{\a\da}} - (\bra{\xi}\rho)^\a\frac{\partial}{\partial \xi^\a},\\
V_{\bar{\rho}} &\equiv -(\bar{\rho}\bar{x})^{\da\a}\frac{\partial}{\partial \bar{x}^{\da\a}} + (p\bar{\rho})_{\a\da}\frac{\partial}{\partial p_{\a\da}} - (\bar{\rho}\sket{\xi})^{\da}\frac{\partial}{\partial \bar{\xi}^{\da}},\\
T_a &\equiv \bar{a}^{\da\a}\frac{\partial}{\partial \bar{x}^{\da\a}},
\end{align}
respectively. We can now compute the corresponding Hamiltonian by considering the action of the symplectic form $\Omega = d\Theta$, viz.
\begin{align}\label{sf:Hamiltonians}
\Omega(R_{\rho},\cdot) = \rd \Tr(\rho J), \qquad \Omega(L_{\bar{\rho}}, \cdot) = d\Tr(\bar{J}\bar{\rho}), \qquad \Omega(T_a,\cdot)=d\Tr(\bar{a}p/2),
\end{align}
where
\begin{align}
J = -\left[\frac{p \bar{x}}{2}  +  i\hbar \frac{p\sket{\xi}\bra{ \xi}}{2m}\frac{\bra{\xi}p
\sket{\xi}}{2ms}\right],\qquad
\bar{J} =  -\left[\frac{\bar{x} p}{2} - i\hbar   \frac{\bra{\xi}p
\sket{\xi}}{2ms}\frac{\sket{\xi}\bra{ \xi} p}{2m} \right].\label{sf:bothJ}
\end{align}
It should be noted that the left and right rotations include left and right dilations. These are obtained by taking $\rho$ and $\br$ proportional to the identity; the generators are
\be
D= -\frac12 \Tr(p\bar{x}),\qquad
R = \hbar \frac{\bra{\xi}p
\sket{\xi}^2}{m^2s}.
\ee
On the other hand, rotations associated with traceless $\rho$ and $\br$ correspond to left and right Lorentz transformations.\\

\indent As noted in \cref{sf:Hamiltonians} the Hamiltonians for right rotations, left rotations, and translations are given by $J$, $\bJ$ and $p/2$, respectively. As such, we can write down the following brackets
\begin{alignat}{4}
\pb{A_\a\,, J_\b^{\pa \gamma}} &=& \delta^\gamma{}_\a A_{\b}, \qquad \pb{\bar{B}_{\da}\,, \bar{J}^{\db}_{\pa \dot{\gamma}}} &=& \delta^\db{}_\da \bar{B}_{\dot{\gamma}}\label{sf:fund1},\\
\pb{A^{\a}\,, J_\b^{\pa \gamma}} &=& -\delta^\a{}_\beta A^{\gamma}, \qquad \pb{\bar{B}^{\da}\,,\bar{J}^{\db}_{\pa \dot{\gamma}}} &=& -\delta^\da{}_{\dot{\gamma}}\bar{B}^{\db}\label{sf:fund2},\\
\pb{x_{\a\da},p_{\b\db}} &=& 2\eps_{\a\b}\eps_{\da\db}, \qquad \pb{\bar{x}^{\da\a}, p_{\b\db}} &=& 2\delta^\a_\b\delta^\da_\db,
\end{alignat}
where $A$ (respectively $\bar{B}$) is any quantity with a single undotted (dotted) index and an unspecified number of dotted (undotted) indices. Note that $J$ commutes with any quantity possessing only undotted indices and vice versa for $\bJ$, furthermore since $p_{\a\da}$, $\xi^\a$, and $\bx^\da$ are invariant under translations they must commute with $p_{\a\da}$. Commutators between the $J$ and $\bJ$ follow from the Jacobi identity
\begin{align}\label{sf:Jbrackets}
\pb{J_\a^{\pa\b}, J_\gamma^{\pa \rho}} = \delta^\rho_\a J_\gamma^{\pa \b} - \delta_\gamma^\b J_\a^{\pa \rho}, \quad \pb{\bJ^\da_{\pa \db}, \bJ^\dg_{\pa \dot{\rho}}} = \delta^\dg_\db \bJ^{\da}_{\pa \dot{\rho}} -\delta^{\da}_{\dot{\rho}} \bJ^\dg_{\pa\db}, \quad \pb{J_\a^{\pa \b}, \bJ^{\da}_{\pa \db}} = 0.
\end{align}
Before we continue, it will be convenient to introduce the null ``position'' vector
\begin{equation}\label{sf:delta}
\bD = \frac{\hbar}{2ms}\frac{\bra{\xi}p\sket{\xi}}{m}\sket{\xi}\bra{ \xi},
\end{equation}
which allow us, c.f. \cref{sf:bothJ}, to parameterize $J$ and $\bar{J}$ as
\begin{align}\label{sf:Jexpand}
J = -\frac{1}{2}p(\bar{x} + i \bD), \qquad \bar{J} = - \frac{1}{2}\left(\bxx - i \bD\right)p.
\end{align}
These expressions can now be inverted to obtain $\bxx$ and $\bD$ in-terms of variables whose Poisson brackets we already know
\begin{align}\label{sf:decomp}
\bar{x} = -\frac{1}{m^2}\left(\bp J + \bar{J}\bp\right), \qquad \bD = \frac{i}{m^2}\left(\bp J - \bar{J} \bp\right).
\end{align}
Using these results we can compute the remaining Poisson brackets, as detailed in Appendix \ref{sf:app:poisson}. One finds that $x$ acts as a generator of translations in momentum space while also rotating the spin variable along an axis determined by $p$, viz
\be
\pb{\bxx^{\da\a},p_{\b\db}} = 2\delta^\a_\b \delta^\da_\db, \qquad \pb{\xi^\a, \bxx^{\db\b}} = \frac{1}{m^2}\bp^{\db\a}\xi^\b.
\ee
Furthermore, the position variable itself is observed to be non-commutative, with the deviation from commutativity being proportional to the spin content. This fundamental modification to the notion of localization is one of the main features of spin, and has been exploited in previous works \cite{Freidel:2007qk}. Explicitly, the $x$ commutation relations read
\be 
\pb{\bxx^{\da\a}, \bxx^{\db\b}} = \frac{i\hbar}{2m s}\frac{\rock}{m^3}\left(\bp^{\da\b}\bx^\db \xi^\a - \bp^{\db\a} \bx^\da \xi^\b\right).
\ee
Last but not least, we witness that the spinor variables behave as creation and annihilation operators: The holomorphic spinors commute with each other, $\pb{\xi^\a, \xi^\b} = 0$, whereas a spinor and its conjugate do not
\be
 \pb{\xi^\a, \bx^\db} = -\frac{is}{\hbar \rock^2}\left(2\rock \bp^{\db\a} - m^2\xi^\a\bx^\db\right).
\ee
This concludes our analysis.

\section{Conclusion}
There are a number of directions for future investigation suggested by the results of this paper. First, can ``half-quantum'' states or the ``Lorentz contraction'' of the four-velocity be observed? Second, it would be interesting to explore the quantum version of this model and see how it compares to the one derived in \cite{rempel_2016}. Specifically, we would like to see how particles of spin half appear in the formalism. Finally, we would like to re-write the interaction vertex between classical spinning particles, originally presented in \cite{rempel_2015}, in-terms of spinors. As we have eliminated all second class constraints this would provide a greater understanding of how the vertex behaves and would allow for a complete path integral formulation of the classical spinning particle.
It would also be important to understand whether we could design an experiment that could distinguish contributions to the quantum phase coming from spin and spacetime motion. Finally, our formalism should naturally allow fir the inclusion of continuous spin particles (CSP's) as a limit where the  mass goes to zero while $ms$ stays fixed. We hope to come back to this in a later publication.

\begin{appendices}
\section{Spinor Formalism}\label{sf:app:spinor}
In this appendix we present a brief review of the spinor helicity formalism, see \cite{Dreiner:2008tw, Martin:1997ns, britto1}. Let $\chi^\a$ be a complex spinor and $\bar{\chi}^{\da} = (\chi^\a)^\dagger$ it's complex conjugate. Indices are raised and lowered with the epsilon tensor $\epsilon_{\alpha\beta}$, which is totally skew symmetric and normalized by $\epsilon_{01}=1$, i.e.
 \begin{align*}
\chi^\a = \eps^{\a\b}\chi_\beta, \quad \chi_\a = \eps_{\a\b}\chi^\b, \quad \bar{\chi}^\da = \eps^{\da\db}\bar{\chi}_{\db}, \quad \bar{\chi}_{\da} = \eps_{\da\db}\bar{\chi}^\db,
\end{align*}
and these quantities are represented as
\begin{align}
\ket{\chi} &= \chi_\a, & \bra{\chi} &= \chi^\a, &
\sket{\chi} &= \bar{\chi}^\da, &  \sbra{\chi} &= \bar{\chi}_\da.
\end{align}
so we see that if $\ket\xi$ is our spinor, the hermitian conjugate spinor is denoted by $\sbra\xi$ as usual while $\sket\xi$ denotes the same spinor but with indices raised. Note that we adopt a convention in which the epsilon tensor satisfies $\eps_{\a\gamma}\eps^{\gamma \beta} = \delta^\b_\a$. 
Contractions between spinors are simply
\begin{align}
\mean{\zeta | \xi} \equiv  \zeta^\a \xi_\a, \qquad \smean{\zeta |\xi} \equiv \bar{\zeta}_\da\bar{\xi}^\da,\qquad   \smean{\zeta |\xi}= -\mean{\zeta | \xi}^*.
\end{align}
Let $(\sig ^a)_{\a\da} = (\Id_{\a\da}, \vec{\sig }_{\a\da})$ be the standard four vector of sigma matrices, and  $(\bar{\sig }^a)^{\da\a} \equiv (\sig ^a)_{\b\db}\eps^{\a\b}\eps^{\da\db}$ the same vector but with indices raised, then the following relations hold 
\begin{align}\label{spinor:sigmacontract}
\mathrm{Tr}(\sig^a\bar{\sig}^b) &= -2\eta^{ab}, \qquad \eta_{ab}(\sig ^a)_{\a\da}(\sig ^b)_{\b\db} = -2 \eps_{\a\b}\eps_{\da\db}.
\end{align}
Generically, a matrix with an overbar is assumed to have upper indices $\bar{M}^{\da\a}$, whereas an unadorned matrix will have lower indices $M_{\a\da}$.  In matrix notation we have that $\bar{M} = \epsilon M^t \epsilon^{-1}$ and  ${\rm{det}}(M)= -\tfrac12\Tr(M\bar{M})$. Multiplication between a matrix and a spinor is denoted by juxtaposition
\begin{align}
M_{\a\da}\bar{\chi}^\da = M\sket{\chi}, \quad \chi^\a M_{\a\da} = \bra{\chi}M, \quad \bar{M}^{\da\a}\chi_\a = \bar{M}\ket{\chi}, \quad \bar{\chi}_\da \bar{M}^{\da\a}  = \sbra{\chi} \bar{M}.
\end{align}
Any vector $p^\mu$ can be represented as a matrix by contracting it with the vector of sigma matrices
\begin{align}
p_\mu = -\frac{1}{2}(\bar{\sig }_\mu)^{\da \a}{p}_{\a \da} \quad \Longleftrightarrow \quad {p}_{\a \da} = p_\mu({\sig }^\mu)_{\a \da}.
\end{align} 
It follows from \cref{spinor:sigmacontract} that 
$p_{\a\da}\bar{p}^{\da\b} = -p^2\delta^\b_\a$ and
$ \bar{p}^{\da\a}p_{\a\db} = -p^2\delta^\da_\db$, so that the inner product of two vectors $p^\mu$ and $q^\mu$ is given by
\begin{align*}
p_\mu q^\mu= -\frac{1}{2}\Tr(p\bar{q}).
\end{align*}
Let $\Lambda^\mu{}_\nu{b}^\nu$ be a Lorentz transformation, then the action of $\Lambda$ on a spinor is represented by matrices $(L_\a{}^\b,\bar{L}^{\da}{}_{\db})$, that is
\begin{align}
\ket{\xi}&\to L\ket{\xi},&\qquad \bra{\xi}&\to \bra{\xi} L^{-1}, \\
\sket{\xi}&\to (L^{-1})^\dagger \sket{\xi}, & \sbra{\xi}&\to \sbra{\xi}L^\dagger.
\end{align} 
The relationship between $\Lambda$ and $(L, \bar{L})$ is obtained through
\be
\bar{L}^{-1} \bar{\sigma}^\mu L = \Lambda^\mu{}_{\nu} \bar{\sigma}^\nu,\qquad
L^{-1} \sigma^\mu \bar{L} =\Lambda^\mu{}_{\nu} {\sigma}^\nu,
\ee
with the $(L, \bar{L})$ satisfying
\be
\bar{L}= (L^{-1})^\dagger,\qquad\epsilon^{\a\a'} L_{\a'}{}^{\b'} \epsilon_{\b'\b}= ([L^{-1}])_\b{}^\a,\qquad
\epsilon_{\da\da'} (\bar{L})^{\da'}{}_{\db'} \epsilon^{\db'\db}= (\bar{L}^{-1})^\db{}_\da= (\bar{L}^\dagger)^\db{}_\da.
\ee
Observe that the contractions we have introduced above are indeed Lorentz invariant. \\
Let us now introduce a structure that involves the contraction of two conjugate spinors along a vector
\be
p^{\da\a} \bar{\zeta}_\da\xi_\a= \sbra{\zeta}\bar{p}\ket{\xi}=  \bra{\xi} p\sket{\zeta}.
\ee
Although this contraction is only invariant under Lorentz transformations that fix $p$, it does have the advantage of defining a hermitian form
\be
\sbra{\zeta}p\ket{\xi}^* = \sbra{\xi}p^\dagger\ket{\zeta} = \sbra{\xi}p\ket{\zeta}.
\ee
Furthermore, if $p$ is a timelike vector $p^2+m^2=0$, this contraction defines a norm $\sbra{\xi}\bar{p}\ket{\xi}$ and in the center of mass frame this 
norm square is simply given by
 $ \pm m (|\xi_0|^2+|\xi_1|^2)$. The sign of the  this 
 scalar product is the sign of the energy $\pm= \mathrm{sign}(p_0)$. \\
 The next thing to consider is the spinorial expression of a bivector. We begin by defining the rotation matrices
 \be
 (\sigma^{\mu\nu})_\a{}^\b \equiv \frac{i}{4}(\sigma^\mu\bar{\sigma}^\nu- \sigma^\nu\bar{\sigma}^\mu)_\a{}^\b,\qquad 
 (\bar{\sigma}^{\mu\nu})^\da{}_\db \equiv \frac{i}{4}(\bar{\sigma}^\mu {\sigma}^\nu - \bar\sigma^\nu{\sigma}^\mu)^\da{}_\db,
 \ee
which can be used to expand the anti-symmetric combination of Pauli matrices
\begin{align}
\sig ^{[\mu}_{\a\da}\sig ^{\nu]}_{\b\db} &= i\eps_{\da\db}(\sig ^{\mu\nu})_{\a\b} - i \eps_{\a\b}(\bar{\sig }^{\mu\nu})_{\da\db},\\
\bar{\sig }^{\da\a}_{[\mu}\bar{\sig }^{\db\b}_{\nu]} &= -i\eps_{\da\db}(\sig ^{\mu\nu})_{\a\b} + i \eps_{\a\b}(\bar{\sig }^{\mu\nu})_{\da\db}.
\end{align}
The rotation matrices possess self-duality properties
 \be
 (*\sigma)^{\mu\nu}= i \sigma^{\mu\nu},
 \qquad
 (*\bar{\sigma})^{\mu\nu}= -i \sigma^{\mu\nu},
 \ee
 where $(*M)^{\mu\nu} = \frac{1}{2}\eps^{\mu\nu\rho\sig}M_{\rho\sig}$ and we have assumed $\eps^{0123} = 1$. A bi-vector  $S_{\mu\nu}$ can be decomposed into self-dual $S_{\a}{}^{\b} = S_{\mu\nu} (\sigma^{\mu\nu})_{\a}{}^{\b}$ and anti self-dual $\bar{S}^{\da}{}_{\db}= S_{\mu\nu} (\bar{\sigma}^{\mu\nu})^\da{}_\db$ parts, specifically
  \bea
 S_{\mu\nu} \sigma^\mu_{\a\da} \sigma^\nu_{\b\db}&=& (iS_{(\a\b)} \epsilon_{\da\db}
 - i\bar{S}_{(\da\db)} \epsilon_{\a\b}), \\
 (*{S})_{\mu\nu} \sigma^\mu_{\a\da} \sigma^\nu_{\b\db}&=& -(S_{(\a\b)} \epsilon_{\da\db}
 +\bar{S}_{(\da\db)} \epsilon_{\a\b}). 
 \eea
 With the spinor indices raised the decomposition is the negative of the one presented above. If the bivector is simple, i.e. $S_{\mu\nu}=(\chi\wedge\pi)_{\mu\nu}$, then we have 
 \bea
S_{\alpha}{}^{\beta}=\frac{i}2(\chi\bar{\pi}-\pi\bar{\chi})_{\a}{}^{\b}
,\quad
\bar{S}^{\da}{}_{\db}= \frac{i}2(\bar{\chi}\pi-\bar{\pi}\chi)^{\da}{}_{\db}
 \eea
 or
 \begin{align}
 S_{\a\b} = -i (\pi\bar{\chi})_{(\a\b)}, \quad \bar{S}_{\da\db} = i(\bar{\chi}\pi)_{(\da\db)}.
 \end{align}
 In other words we can express the matrix product of two vectors as
 \be
 (\chi\bar{\pi})_{\alpha}{}^{\beta} = -(\chi_\mu\pi^\mu) \delta_\a^\b -i(\chi\wedge\pi)_{\alpha}{}^{\beta},\qquad
   (\bar{\chi}{\pi})^{\da}{}_{\db} = -(\chi_\mu\pi^\mu) \delta_\db^\da -i(\chi\wedge\pi)^{\da}{}_{\db}.
 \ee
The matrix corresponding to a vector $p_\mu$ can be expressed explicitly as
 \be
 p_{\a\da}= \left( \begin{array}{cc}
 (p_0 + p_3) & (p_1-ip_2)  \\
(p_1+ip_2)  &(p_0 - p_3) \\
  \end{array} \right),\qquad 
  \bar{p}^{\da\a} = \left( \begin{array}{cc}
 (p_0 - p_3) & -(p_1-ip_2)  \\
-(p_1+ip_2)  &(p_0 + p_3) \\
  \end{array} \right).
 \ee
We see that the bar operator corresponds to parity reversal, that is, if we denote the parity transformed vector  $\tilde{p}_\mu\equiv (p_0,-p_i)$ then  $\bar{p}=\tilde{p}$ as matrices. We also find that
\be
(\chi \bar{\pi})_{\a}{}^\b = (\chi_\mu \pi^\mu) 1 +i
 \left( \begin{array}{cc}
 J_3+iK_3 & (J_1+iK_1) -i(J_2+iK_2)  \\
(J_1+iK_1) +i(J_2+iK_2)  &-(J_3+iK_3 ) \\
  \end{array} \right) 
\ee
where we have defined 
\be
K_i = (\chi \wedge \pi)_{i0},\qquad J_i =\epsilon_{ijk}(\chi \wedge \pi)^{jk},
\ee
as ``boost'' and ``rotation'' generators respectively.

\section{First Order Equations of Motion}\label{sf:app:first}
In this Appendix we verify that the spin state does not evolve during classical motion. The equations of motion associated with the Lagrangian \cref{firstorder} are given by
\begin{align}
\frac{d}{dt}P &= 0,\\
\frac{d}{dt}\left(\bra{\xi}P\right) &= -\left(\bra{\dxi} +2i\dot{\phi}\bra{\xi} \right) P.
\end{align}
where we have defined
\be
P:= \frac1{N} \left(\dot{x} +  \theta(e^{i\phi}\xi)\right).
\ee
The first of these implies that $P_{\a\da}$ is constant, and inserting this result into the second equation gives
\begin{align}\label{sf:full_classical}
\left(\bra{\dxi} + i\dot{\phi}\bra{\xi}\right)C = 0 \qquad \Longrightarrow \qquad \frac{d}{dt}\left(e^{i\phi}\bra{\xi}\right) = 0.
\end{align}
It follows that $e^{i\phi}\xi$ is a constant of motion and so $\theta(e^{i\phi}\xi)$ vanishes on-shell.

\section{Second Order Equations of Motion}\label{sf:app:second}
In what follows we will derive the equations of motion associated with the second order action \cref{sf:Actionf}. We begin by defining the momentum $p_\mu$ conjugate to $x_\mu$ in the usual manner $p_\mu = \delta S/ \delta \dot{x}^\mu$, then the equation of motion for $x$ is determined by conservation of momenta $\dot{p} = 0$. As an important aside, the relationship between $p_{\a\da}$ and $\delta S/\delta \bar{x}^{\da\a}$ isn't quite as expected, specifically
\begin{align}
p_{\a\da} = (\sig ^\mu)_{\a\da}p_\mu = (\sig ^\mu_{\a\da})\frac{\delta L}{\delta \dot{x}^\mu} = (\sig ^\mu_{\a\da})\frac{\delta L}{\delta \dot{\bar{x}}^{\db\b}}\frac{\partial \dot{\bar{x}}^{\db\b}}{\partial \dot{x}^\mu} = -2\frac{\delta L}{\delta \dot{\bar{x}}^{\da\a}}.
\end{align}
Recalling the definitions of $\mf{a}$ and $\mf{b}$ from \cref{mfaANDmfB} we find that the momenta $p_{\a\da}$ is given by
\begin{align}\label{sf:momentum}
-\frac{1}{2}p &= m\frac{\dot{x}}{2\abs{\dot{x}}}\sqrt{1 - \abs{\mf{b}}^2} - \frac{i\hbar s}{\rockX}\Big(\ket{\dot{\rho}}\sbra{\rho} - \ket{\rho}\sbra{\dot{\rho}}\Big) \\
&\qquad - \frac{2\hbar s}{\rockX}\mr{Im}(\mf{a}) \ket{\rho}\sbra{\rho}
+ \frac{m\abs{\mf{b}}^2}{\sqrt{1 - \abs{\mf{b}}^2}}\frac{\ket{\rho}\abs{\dot{x}}\sbra{\rho}}{\rockX}\nonumber.
\end{align}
It follows from this lengthy expression that
\begin{align*}
\dot{x}\sket{\rho} &= -\frac{\abs{\dot{x}}}{\sqrt{1-\abs{\mf{b}}^2}}\Big(\hat{p}\sket{\rho} + i\mf{b}^*\ket{\rho}\Big),
\end{align*}
where $\hat{p}$ is the unit momenta $\hat{p} = p/m$. The spin equations of motion can now be written in matrix form as
\begin{align*}
\partial_\tau\left(
\begin{array}{c}
\ket{\rho}\\
\hat{p}\sket{\rho}
\end{array}
\right)
=
\left(
\begin{array}{cc}
\mf{a} - \frac{im}{2\hbar s}\frac{\abs{\mf{b}}^2}{ \sqrt{1-\abs{\mf{b}}^2}}\abs{\dot{x}} & -\frac{m\mf{b}}{2\hbar s}\frac{ \abs{\dot{x}}}{\sqrt{1-\abs{\mf{b}}^2}}\\
\frac{m\mf{b}^*}{2\hbar s}\frac{\abs{\dot{x}}}{\sqrt{1 - \abs{\mf{b}}^2}} &   \mf{a}^* + \frac{im}{2\hbar s}\frac{\abs{\mf{b}}^2}{\sqrt{1 - \abs{\mf{b}}^2}}\abs{\dot{x}}
\end{array}
\right)
\left(
\begin{array}{c}
\ket{\rho}\\
\hat{p}\sket{\rho}
\end{array}
\right).
\end{align*}
To simplify the presentation we introduce the notation
\begin{gather*}
\boldsymbol{\rho} = \left(\begin{array}{c}
\ket{\rho}\\
\hat{p}\sket{\rho}
\end{array}
\right),
 \quad a_0 = \mr{Re}(\mf{a}), \quad a_1 = -\frac{m}{2\hbar s}\frac{ \abs{\dot{x}}}{\sqrt{1-\abs{\mf{b}}^2}}\mr{Im}(\mf{b}),\\
a_2 = -\frac{m}{2\hbar s}\frac{ \abs{\dot{x}}}{\sqrt{1-\abs{\mf{b}}^2}}\mr{Re}(\mf{b}), \quad a_3 = \mr{Im}(\mf{a}) - \frac{m}{2\hbar s}\frac{\abs{\mf{b}}^2}{ \sqrt{1-\abs{\mf{b}}^2}}\abs{\dot{x}},
\end{gather*}
and so the spin equations of motion become
\begin{align}
\partial_\tau \boldsymbol{\rho} &= 
\left(
\begin{array}{cc}
a_0 + i a_3 & ia_1 + a_2\\
ia_1 - a_2 & a_0 - ia_3
\end{array}
\right)\boldsymbol{\rho},\\
&= \left(a_0\Id + i\vec{a}\cdot \vec{\sig }\right)\boldsymbol{\rho},\label{sf:boldrho}
\end{align}
where $\vec{a} = (a_1,a_2, a_3)$. To solve this equation we introduce a vector $\bsy{\xi}$ which satisfies
\begin{align}
\bsy{\rho} = e^{\int_0^\tau a_0(t)dt}\bsy{\xi}.
\end{align}
\Cref{sf:boldrho} then implies $\partial_\tau \bsy{\xi} = i\vec{a}\cdot\vec{\sig} \bsy{\xi}$, which has the formal solution
\begin{align}\label{sf:formalSol}
\bsy{\xi} = \mr{T}\exp\left(i\vec{\sig }\cdot\int_0^\tau \vec{a}(t)dt\right)\bsy{\xi}(0).
\end{align}
It is only in the special case where at most one component of $\vec{a}$ is non-zero that we could obtain an explicit expression for $\bsy{\xi}$.

\section{Poisson Brackets}\label{sf:app:poisson}
This Appendix provides some additional details regarding the computation of the Poisson algebra for the spinor variables. \Crefrange{sf:fund1}{sf:Jbrackets} together with the expression for $x$ and $\Delta $ in terms of 
$J$ given by (\ref{sf:decomp}) imply that
\begin{alignat}{4}
\pb{(\bxx + i \bD)^{\da\a},(\bxx + i \bD)^{\db\b}} &=& 0, \qquad \pb{(\bxx + i \bD)^{\da\a}, (\bxx - i \bD)^{\db\b}} &=& \frac{4i}{m^2}\bp^{\da\b}\bD^{\db\a},\\
\pb{(\bxx - i \bD)^{\da\a}, (\bxx - i \bD)^{\db\b}} &=& 0, \qquad \pb{(\bxx - i \bD)^{\da\a}, (\bxx + i \bD)^{\db\b}} &=& -\frac{4i}{m^2}\bp^{\db\a}\bD^{\da\b},
\end{alignat}
which combine to give
\begin{gather}
\pb{\bxx^{\da\a}, \bxx^{\db\b}} = \pb{\bD^{\da\a}, \bD^{\db\b}} = \frac{i}{m^2}\Big(\bp^{\da\b}\bD^{\db\a} - \bp^{\db\a} \bD^{\da\b}\Big),\label{sf:xxdd}\\
\pb{\bxx^{\da\a}, \bD^{\db\b}} = - \frac{1}{m^2}\Big(\bp^{\da\b}\bD^{\db\a} + \bp^{\db\a} \bD^{\da\b}\Big).
\end{gather}
Noting that the brackets in \cref{sf:xxdd} are anti-symmetric under interchange of $(\a,\da)$ with $(\b, \db)$ allows us to re-write them in a more revealing form
\begin{align}
\pb{\bxx^{\da\a}, \bxx^{\db\b}} = \pb{\bD^{\da\a}, \bD^{\db\b}} = \frac{i}{m^2}\left[\eps^{\da\db}(p\bD)^{(\a\b)}  - \eps^{\a\b}(\bD p)^{(\da\db)}\right].
\end{align}
A further application of \crefrange{sf:fund1}{sf:fund2} to $\xi$ and $\bx$ in conjunction with the decomposition \eqref{sf:decomp} yields
\begin{alignat}{4}
\pb{\xi^\a, \bxx^{\db\b}} &=& \frac{1}{m^2}\bp^{\db\a}\xi^\b, \qquad \pb{\bx^{\da}, \bxx^{\db\b}} &=& \frac{1}{m^2}\bp^{\da\b}\bx^{\db},\\
\pb{\xi^\a, \bD^{\db\b}} &=& -\frac{i}{m^2}\bp^{\db\a}\xi^\b, \qquad \pb{\bx^{\da}, \bD^{\db\b}} &=& \frac{i}{m^2}\bp^{\da\b}\bx^{\db}\label{sf:xiDelta}.
\end{alignat}
It remains to compute the brackets between $\xi$ and $\bx$. We 
begin by substituting the definition of $\Delta$, see \cref{sf:delta}, into \cref{sf:xiDelta}, whence
\begin{align}\label{sf:deltaExpand}
\frac{\hbar}{2m^2s}\rock\left(\xi^\b\pb{\xi^\a, \bx^{\db}} + \bx^{\db}\pb{\xi^\a, \xi^\b}\right) = -\frac{i}{m^2}\bp^{\db\a}\xi^\b + \frac{i}{2\rock}\xi^\a\xi^\b\bx^\db.
\end{align}
Contract either side with $\xi_\b$ to obtain $\pb{\xi^\a, \xi^\b}\xi_\b = 0$ which, by virtue of the anti--symmetry of the bracket, implies
\begin{align}
\pb{\xi^\a, \xi^\b} = 0.
\end{align}
Upon substituting the above result into \cref{sf:deltaExpand} and contracting with $(p\sket{\xi})_\b$ we obtain 
\begin{align}
\pb{\xi^\a, \bx^\db} = -\frac{is}{\hbar \rock^2}\left(2\rock \bp^{\db\a} - m^2\xi^\a\bx^\db\right).
\end{align}
Similar results hold for $\bx$, in particular 
\begin{align}
\pb{\bx^\da, \bx^{\db}} = 0 , \qquad \pb{\bx^\da, \xi^\b} = \frac{is}{\hbar \rock^2}\left(2\rock \bp^{\da\b} - m^2\xi^\b\bx^\da\right).
\end{align}
\end{appendices}

\bibliography{spinor_ref}{}

\begin{thebibliography}{31}%
\makeatletter
\providecommand \@ifxundefined [1]{%
 \@ifx{#1\undefined}
}%
\providecommand \@ifnum [1]{%
 \ifnum #1\expandafter \@firstoftwo
 \else \expandafter \@secondoftwo
 \fi
}%
\providecommand \@ifx [1]{%
 \ifx #1\expandafter \@firstoftwo
 \else \expandafter \@secondoftwo
 \fi
}%
\providecommand \natexlab [1]{#1}%
\providecommand \enquote  [1]{``#1''}%
\providecommand \bibnamefont  [1]{#1}%
\providecommand \bibfnamefont [1]{#1}%
\providecommand \citenamefont [1]{#1}%
\providecommand \href@noop [0]{\@secondoftwo}%
\providecommand \href [0]{\begingroup \@sanitize@url \@href}%
\providecommand \@href[1]{\@@startlink{#1}\@@href}%
\providecommand \@@href[1]{\endgroup#1\@@endlink}%
\providecommand \@sanitize@url [0]{\catcode `\\12\catcode `\$12\catcode
  `\&12\catcode `\#12\catcode `\^12\catcode `\_12\catcode `\%12\relax}%
\providecommand \@@startlink[1]{}%
\providecommand \@@endlink[0]{}%
\providecommand \url  [0]{\begingroup\@sanitize@url \@url }%
\providecommand \@url [1]{\endgroup\@href {#1}{\urlprefix }}%
\providecommand \urlprefix  [0]{URL }%
\providecommand \Eprint [0]{\href }%
\providecommand \doibase [0]{http://dx.doi.org/}%
\providecommand \selectlanguage [0]{\@gobble}%
\providecommand \bibinfo  [0]{\@secondoftwo}%
\providecommand \bibfield  [0]{\@secondoftwo}%
\providecommand \translation [1]{[#1]}%
\providecommand \BibitemOpen [0]{}%
\providecommand \bibitemStop [0]{}%
\providecommand \bibitemNoStop [0]{.\EOS\space}%
\providecommand \EOS [0]{\spacefactor3000\relax}%
\providecommand \BibitemShut  [1]{\csname bibitem#1\endcsname}%
\let\auto@bib@innerbib\@empty
\bibitem [{\citenamefont {Schr{\"o}dinger}(1930)}]{schrodinger_1930}%
  \BibitemOpen
  \bibfield  {author} {\bibinfo {author} {\bibfnamefont {E.}~\bibnamefont
  {Schr{\"o}dinger}},\ }\href@noop {} {\bibfield  {journal} {\bibinfo
  {journal} {Berliner Ber.}\ ,\ \bibinfo {pages} {418}} (\bibinfo {year}
  {1930})}\BibitemShut {NoStop}%
\bibitem [{\citenamefont {Rempel}\ and\ \citenamefont
  {Freidel}(2016{\natexlab{a}})}]{rempel_2015}%
  \BibitemOpen
  \bibfield  {author} {\bibinfo {author} {\bibfnamefont {T.}~\bibnamefont
  {Rempel}}\ and\ \bibinfo {author} {\bibfnamefont {L.}~\bibnamefont
  {Freidel}},\ }\href {\doibase 10.1103/PhysRevD.94.044011} {\bibfield
  {journal} {\bibinfo  {journal} {Phys. Rev.}\ }\textbf {\bibinfo {volume}
  {D94}},\ \bibinfo {pages} {044011} (\bibinfo {year} {2016}{\natexlab{a}})},\
  \Eprint {http://arxiv.org/abs/1507.05826} {arXiv:1507.05826 [hep-th]}
  \BibitemShut {NoStop}%
\bibitem [{\citenamefont {Frydryszak}(1996)}]{Frydryszak:1996mu}%
  \BibitemOpen
  \bibfield  {author} {\bibinfo {author} {\bibfnamefont {A.}~\bibnamefont
  {Frydryszak}},\ }\href@noop {} {\  (\bibinfo {year} {1996})},\ \Eprint
  {http://arxiv.org/abs/hep-th/9601020} {arXiv:hep-th/9601020 [hep-th]}
  \BibitemShut {NoStop}%
\bibitem [{\citenamefont {Hanson}\ and\ \citenamefont
  {Regge}(1974)}]{hanson_1974}%
  \BibitemOpen
  \bibfield  {author} {\bibinfo {author} {\bibfnamefont {A.~J.}\ \bibnamefont
  {Hanson}}\ and\ \bibinfo {author} {\bibfnamefont {T.}~\bibnamefont {Regge}},\
  }\href {\doibase 10.1016/0003-4916(74)90046-3} {\bibfield  {journal}
  {\bibinfo  {journal} {Annals Phys.}\ }\textbf {\bibinfo {volume} {87}},\
  \bibinfo {pages} {498} (\bibinfo {year} {1974})}\BibitemShut {NoStop}%
\bibitem [{\citenamefont {Balachandran}\ \emph {et~al.}(1977)\citenamefont
  {Balachandran}, \citenamefont {Salomonson}, \citenamefont {Skagerstam},\ and\
  \citenamefont {Winnberg}}]{balachandran_1976}%
  \BibitemOpen
  \bibfield  {author} {\bibinfo {author} {\bibfnamefont {A.}~\bibnamefont
  {Balachandran}}, \bibinfo {author} {\bibfnamefont {P.}~\bibnamefont
  {Salomonson}}, \bibinfo {author} {\bibfnamefont {B.-S.}\ \bibnamefont
  {Skagerstam}}, \ and\ \bibinfo {author} {\bibfnamefont {J.-O.}\ \bibnamefont
  {Winnberg}},\ }\href {\doibase 10.1103/PhysRevD.15.2308} {\bibfield
  {journal} {\bibinfo  {journal} {Phys.Rev.}\ }\textbf {\bibinfo {volume}
  {D15}},\ \bibinfo {pages} {2308} (\bibinfo {year} {1977})}\BibitemShut
  {NoStop}%
\bibitem [{\citenamefont {Balachandran}\ \emph {et~al.}(1980)\citenamefont
  {Balachandran}, \citenamefont {Marmo}, \citenamefont {Skagerstam},\ and\
  \citenamefont {Stern}}]{balachandran_1979}%
  \BibitemOpen
  \bibfield  {author} {\bibinfo {author} {\bibfnamefont {A.}~\bibnamefont
  {Balachandran}}, \bibinfo {author} {\bibfnamefont {G.}~\bibnamefont {Marmo}},
  \bibinfo {author} {\bibfnamefont {B.}~\bibnamefont {Skagerstam}}, \ and\
  \bibinfo {author} {\bibfnamefont {A.}~\bibnamefont {Stern}},\ }\href
  {\doibase 10.1016/0370-2693(80)90009-X} {\bibfield  {journal} {\bibinfo
  {journal} {Phys.Lett.}\ }\textbf {\bibinfo {volume} {B89}},\ \bibinfo {pages}
  {199} (\bibinfo {year} {1980})}\BibitemShut {NoStop}%
\bibitem [{\citenamefont {Balachandran}\ \emph {et~al.}(1982)\citenamefont
  {Balachandran}, \citenamefont {Marmo}, \citenamefont {Mukunda}, \citenamefont
  {Nilsson}, \citenamefont {Simoni} \emph {et~al.}}]{balachandran_1981}%
  \BibitemOpen
  \bibfield  {author} {\bibinfo {author} {\bibfnamefont {A.}~\bibnamefont
  {Balachandran}}, \bibinfo {author} {\bibfnamefont {G.}~\bibnamefont {Marmo}},
  \bibinfo {author} {\bibfnamefont {N.}~\bibnamefont {Mukunda}}, \bibinfo
  {author} {\bibfnamefont {J.}~\bibnamefont {Nilsson}}, \bibinfo {author}
  {\bibfnamefont {A.}~\bibnamefont {Simoni}},  \emph {et~al.},\ }\href
  {\doibase 10.1007/BF02816669} {\bibfield  {journal} {\bibinfo  {journal}
  {Nuovo Cim.}\ }\textbf {\bibinfo {volume} {A67}},\ \bibinfo {pages} {121}
  (\bibinfo {year} {1982})}\BibitemShut {NoStop}%
\bibitem [{\citenamefont {Kirillov}(1976)}]{kirillov_1976}%
  \BibitemOpen
  \bibfield  {author} {\bibinfo {author} {\bibfnamefont {A.}~\bibnamefont
  {Kirillov}},\ }\href@noop {} {\emph {\bibinfo {title} {{Elements of the
  Theory of Representations}}}}\ (\bibinfo  {publisher} {Springer--Verlag.},\
  \bibinfo {year} {1976})\BibitemShut {NoStop}%
\bibitem [{\citenamefont {Kostant}(1970)}]{kostant_1970}%
  \BibitemOpen
  \bibfield  {author} {\bibinfo {author} {\bibfnamefont {B.}~\bibnamefont
  {Kostant}},\ }in\ \href {\doibase 10.1007/BFb0079068} {\emph {\bibinfo
  {booktitle} {Lectures in Modern Analysis and Applications III}}},\ \bibinfo
  {series} {Lecture Notes in Mathematics}, Vol.\ \bibinfo {volume} {170},\
  \bibinfo {editor} {edited by\ \bibinfo {editor} {\bibfnamefont
  {C.}~\bibnamefont {Taam}}}\ (\bibinfo  {publisher} {Springer Berlin
  Heidelberg},\ \bibinfo {year} {1970})\ pp.\ \bibinfo {pages}
  {87--208}\BibitemShut {NoStop}%
\bibitem [{\citenamefont {Souriau}(1970)}]{souriau_1970}%
  \BibitemOpen
  \bibfield  {author} {\bibinfo {author} {\bibfnamefont {J.}~\bibnamefont
  {Souriau}},\ }\href@noop {} {\emph {\bibinfo {title} {{Structure des Systemes
  Dynamiques}}}}\ (\bibinfo  {publisher} {Dunod},\ \bibinfo {year}
  {1970})\BibitemShut {NoStop}%
\bibitem [{\citenamefont {Souriau}(1997)}]{souriau_1997}%
  \BibitemOpen
  \bibfield  {author} {\bibinfo {author} {\bibfnamefont {J.}~\bibnamefont
  {Souriau}},\ }\href@noop {} {\emph {\bibinfo {title} {{Structure of Dynamical
  Systems: A symplectic view of physics}}}}\ (\bibinfo  {publisher} {Birkhauser
  Boston.},\ \bibinfo {year} {1997})\BibitemShut {NoStop}%
\bibitem [{\citenamefont {Wiegmann}(1989)}]{wiegmann_1989B}%
  \BibitemOpen
  \bibfield  {author} {\bibinfo {author} {\bibfnamefont {P.}~\bibnamefont
  {Wiegmann}},\ }\href {\doibase 10.1016/0550-3213(89)90144-2} {\bibfield
  {journal} {\bibinfo  {journal} {Nucl.Phys.}\ }\textbf {\bibinfo {volume}
  {B323}},\ \bibinfo {pages} {311} (\bibinfo {year} {1989})}\BibitemShut
  {NoStop}%
\bibitem [{\citenamefont {Zakrzewski}(1995)}]{Zakrzewski:1994bd}%
  \BibitemOpen
  \bibfield  {author} {\bibinfo {author} {\bibfnamefont {S.}~\bibnamefont
  {Zakrzewski}},\ }\href {\doibase 10.1088/0305-4470/28/24/028} {\bibfield
  {journal} {\bibinfo  {journal} {J. Phys.}\ }\textbf {\bibinfo {volume}
  {A28}},\ \bibinfo {pages} {7347} (\bibinfo {year} {1995})},\ \Eprint
  {http://arxiv.org/abs/hep-th/9412100} {arXiv:hep-th/9412100 [hep-th]}
  \BibitemShut {NoStop}%
\bibitem [{\citenamefont {Deriglazov}(2013)}]{Deriglazov:2012kw}%
  \BibitemOpen
  \bibfield  {author} {\bibinfo {author} {\bibfnamefont {A.~A.}\ \bibnamefont
  {Deriglazov}},\ }\href {\doibase 10.1142/S0217732312502343} {\bibfield
  {journal} {\bibinfo  {journal} {Mod. Phys. Lett.}\ }\textbf {\bibinfo
  {volume} {A28}},\ \bibinfo {pages} {1250234} (\bibinfo {year} {2013})},\
  \Eprint {http://arxiv.org/abs/1204.2494} {arXiv:1204.2494 [hep-th]}
  \BibitemShut {NoStop}%
\bibitem [{\citenamefont {Plyushchay}(1990)}]{Plyushchay:1990cz}%
  \BibitemOpen
  \bibfield  {author} {\bibinfo {author} {\bibfnamefont {M.~S.}\ \bibnamefont
  {Plyushchay}},\ }\href {\doibase 10.1016/0370-2693(90)90296-I} {\bibfield
  {journal} {\bibinfo  {journal} {Phys. Lett.}\ }\textbf {\bibinfo {volume}
  {B248}},\ \bibinfo {pages} {299} (\bibinfo {year} {1990})}\BibitemShut
  {NoStop}%
\bibitem [{\citenamefont {Penrose}\ and\ \citenamefont
  {Rindler}(2011)}]{Penrose:1987uia}%
  \BibitemOpen
  \bibfield  {author} {\bibinfo {author} {\bibfnamefont {R.}~\bibnamefont
  {Penrose}}\ and\ \bibinfo {author} {\bibfnamefont {W.}~\bibnamefont
  {Rindler}},\ }\href
  {http://www.cambridge.org/mw/academic/subjects/physics/theoretical-physics-and-mathematical-physics/spinors-and-space-time-volume-1?format=AR}
  {\emph {\bibinfo {title} {{Spinors and Space-Time}}}},\ Cambridge Monographs
  on Mathematical Physics\ (\bibinfo  {publisher} {Cambridge Univ. Press},\
  \bibinfo {address} {Cambridge, UK},\ \bibinfo {year} {2011})\BibitemShut
  {NoStop}%
\bibitem [{\citenamefont {Bengtsson}\ \emph {et~al.}(1987)\citenamefont
  {Bengtsson}, \citenamefont {Bengtsson}, \citenamefont {Cederwall},\ and\
  \citenamefont {Linden}}]{Bengtsson:1987ap}%
  \BibitemOpen
  \bibfield  {author} {\bibinfo {author} {\bibfnamefont {A.~K.~H.}\
  \bibnamefont {Bengtsson}}, \bibinfo {author} {\bibfnamefont {I.}~\bibnamefont
  {Bengtsson}}, \bibinfo {author} {\bibfnamefont {M.}~\bibnamefont
  {Cederwall}}, \ and\ \bibinfo {author} {\bibfnamefont {N.}~\bibnamefont
  {Linden}},\ }\href {\doibase 10.1103/PhysRevD.36.1766} {\bibfield  {journal}
  {\bibinfo  {journal} {Phys. Rev.}\ }\textbf {\bibinfo {volume} {D36}},\
  \bibinfo {pages} {1766} (\bibinfo {year} {1987})}\BibitemShut {NoStop}%
\bibitem [{\citenamefont {Kuzenko}\ \emph {et~al.}(1995)\citenamefont
  {Kuzenko}, \citenamefont {Lyakhovich},\ and\ \citenamefont
  {Segal}}]{lyakhovich_1994}%
  \BibitemOpen
  \bibfield  {author} {\bibinfo {author} {\bibfnamefont {S.}~\bibnamefont
  {Kuzenko}}, \bibinfo {author} {\bibfnamefont {S.}~\bibnamefont {Lyakhovich}},
  \ and\ \bibinfo {author} {\bibfnamefont {A.~Y.}\ \bibnamefont {Segal}},\
  }\href {\doibase 10.1142/S0217751X95000735} {\bibfield  {journal} {\bibinfo
  {journal} {Int.J.Mod.Phys.}\ }\textbf {\bibinfo {volume} {A10}},\ \bibinfo
  {pages} {1529} (\bibinfo {year} {1995})},\ \Eprint
  {http://arxiv.org/abs/hep-th/9403196} {arXiv:hep-th/9403196 [hep-th]}
  \BibitemShut {NoStop}%
\bibitem [{\citenamefont {Kassandrov}\ \emph {et~al.}(2009)\citenamefont
  {Kassandrov}, \citenamefont {Markova}, \citenamefont {Schaefer},\ and\
  \citenamefont {Wipf}}]{Kassandrov:2009jd}%
  \BibitemOpen
  \bibfield  {author} {\bibinfo {author} {\bibfnamefont {V.}~\bibnamefont
  {Kassandrov}}, \bibinfo {author} {\bibfnamefont {N.}~\bibnamefont {Markova}},
  \bibinfo {author} {\bibfnamefont {G.}~\bibnamefont {Schaefer}}, \ and\
  \bibinfo {author} {\bibfnamefont {A.}~\bibnamefont {Wipf}},\ }\href {\doibase
  10.1088/1751-8113/42/31/315204} {\bibfield  {journal} {\bibinfo  {journal}
  {J. Phys.}\ }\textbf {\bibinfo {volume} {A42}},\ \bibinfo {pages} {315204}
  (\bibinfo {year} {2009})},\ \Eprint {http://arxiv.org/abs/0902.3688}
  {arXiv:0902.3688 [hep-th]} \BibitemShut {NoStop}%
\bibitem [{\citenamefont {Barut}\ and\ \citenamefont
  {Zanghi}(1984)}]{barut_1984B}%
  \BibitemOpen
  \bibfield  {author} {\bibinfo {author} {\bibfnamefont {A.}~\bibnamefont
  {Barut}}\ and\ \bibinfo {author} {\bibfnamefont {N.}~\bibnamefont {Zanghi}},\
  }\href {\doibase 10.1103/PhysRevLett.52.2009} {\bibfield  {journal} {\bibinfo
   {journal} {Phys. Rev. Lett.}\ }\textbf {\bibinfo {volume} {52}},\ \bibinfo
  {pages} {2009} (\bibinfo {year} {1984})}\BibitemShut {NoStop}%
\bibitem [{\citenamefont {Pavsic}\ \emph {et~al.}(1993)\citenamefont {Pavsic},
  \citenamefont {Recami}, \citenamefont {Rodrigues}, \citenamefont
  {Maccarrone}, \citenamefont {Raciti} \emph {et~al.}}]{pavsic_1992}%
  \BibitemOpen
  \bibfield  {author} {\bibinfo {author} {\bibfnamefont {M.}~\bibnamefont
  {Pavsic}}, \bibinfo {author} {\bibfnamefont {E.}~\bibnamefont {Recami}},
  \bibinfo {author} {\bibfnamefont {J.}~\bibnamefont {Rodrigues}, \bibfnamefont
  {Waldyr~A.}}, \bibinfo {author} {\bibfnamefont {G.~D.}\ \bibnamefont
  {Maccarrone}}, \bibinfo {author} {\bibfnamefont {F.}~\bibnamefont {Raciti}},
  \emph {et~al.},\ }\href {\doibase 10.1016/0370-2693(93)91543-V} {\bibfield
  {journal} {\bibinfo  {journal} {Phys.Lett.}\ }\textbf {\bibinfo {volume}
  {B318}},\ \bibinfo {pages} {481} (\bibinfo {year} {1993})}\BibitemShut
  {NoStop}%
\bibitem [{\citenamefont {Lyakhovich}\ \emph {et~al.}(1996)\citenamefont
  {Lyakhovich}, \citenamefont {Segal},\ and\ \citenamefont
  {Sharapov}}]{lyakhovich_1996}%
  \BibitemOpen
  \bibfield  {author} {\bibinfo {author} {\bibfnamefont {S.}~\bibnamefont
  {Lyakhovich}}, \bibinfo {author} {\bibfnamefont {A.~Y.}\ \bibnamefont
  {Segal}}, \ and\ \bibinfo {author} {\bibfnamefont {A.}~\bibnamefont
  {Sharapov}},\ }\href {\doibase 10.1103/PhysRevD.54.5223} {\bibfield
  {journal} {\bibinfo  {journal} {Phys.Rev.}\ }\textbf {\bibinfo {volume}
  {D54}},\ \bibinfo {pages} {5223} (\bibinfo {year} {1996})},\ \Eprint
  {http://arxiv.org/abs/hep-th/9603174} {arXiv:hep-th/9603174 [hep-th]}
  \BibitemShut {NoStop}%
\bibitem [{\citenamefont {Lyakhovich}\ \emph {et~al.}(1999)\citenamefont
  {Lyakhovich}, \citenamefont {Sharapov},\ and\ \citenamefont
  {Shekhter}}]{lyakhovich_1998A}%
  \BibitemOpen
  \bibfield  {author} {\bibinfo {author} {\bibfnamefont {S.}~\bibnamefont
  {Lyakhovich}}, \bibinfo {author} {\bibfnamefont {A.}~\bibnamefont
  {Sharapov}}, \ and\ \bibinfo {author} {\bibfnamefont {K.}~\bibnamefont
  {Shekhter}},\ }\href {\doibase 10.1016/S0550-3213(98)00617-8} {\bibfield
  {journal} {\bibinfo  {journal} {Nucl.Phys.}\ }\textbf {\bibinfo {volume}
  {B537}},\ \bibinfo {pages} {640} (\bibinfo {year} {1999})},\ \Eprint
  {http://arxiv.org/abs/hep-th/9805020} {arXiv:hep-th/9805020 [hep-th]}
  \BibitemShut {NoStop}%
\bibitem [{\citenamefont {Lyakhovich}\ \emph {et~al.}(1998)\citenamefont
  {Lyakhovich}, \citenamefont {Sharapov},\ and\ \citenamefont
  {Shekhter}}]{lyakhovich_1998B}%
  \BibitemOpen
  \bibfield  {author} {\bibinfo {author} {\bibfnamefont {S.}~\bibnamefont
  {Lyakhovich}}, \bibinfo {author} {\bibfnamefont {A.}~\bibnamefont
  {Sharapov}}, \ and\ \bibinfo {author} {\bibfnamefont {K.}~\bibnamefont
  {Shekhter}},\ }\href@noop {} {\  (\bibinfo {year} {1998})},\ \Eprint
  {http://arxiv.org/abs/hep-th/9811003} {arXiv:hep-th/9811003 [hep-th]}
  \BibitemShut {NoStop}%
\bibitem [{\citenamefont {Rempel}\ and\ \citenamefont
  {Freidel}(2016{\natexlab{b}})}]{rempel_2016}%
  \BibitemOpen
  \bibfield  {author} {\bibinfo {author} {\bibfnamefont {T.}~\bibnamefont
  {Rempel}}\ and\ \bibinfo {author} {\bibfnamefont {L.}~\bibnamefont
  {Freidel}},\ }\href@noop {} {\  (\bibinfo {year} {2016}{\natexlab{b}})},\
  \Eprint {http://arxiv.org/abs/1609.09110} {arXiv:1609.09110 [hep-th]}
  \BibitemShut {NoStop}%
\bibitem [{\citenamefont {Freidel}\ \emph {et~al.}(2007)\citenamefont
  {Freidel}, \citenamefont {Girelli},\ and\ \citenamefont
  {Livine}}]{Freidel:2007qk}%
  \BibitemOpen
  \bibfield  {author} {\bibinfo {author} {\bibfnamefont {L.}~\bibnamefont
  {Freidel}}, \bibinfo {author} {\bibfnamefont {F.}~\bibnamefont {Girelli}}, \
  and\ \bibinfo {author} {\bibfnamefont {E.~R.}\ \bibnamefont {Livine}},\
  }\href {\doibase 10.1103/PhysRevD.75.105016} {\bibfield  {journal} {\bibinfo
  {journal} {Phys. Rev.}\ }\textbf {\bibinfo {volume} {D75}},\ \bibinfo {pages}
  {105016} (\bibinfo {year} {2007})},\ \Eprint
  {http://arxiv.org/abs/hep-th/0701113} {arXiv:hep-th/0701113 [hep-th]}
  \BibitemShut {NoStop}%
\bibitem [{\citenamefont {Das}\ and\ \citenamefont {Ghosh}(2009)}]{Das:2009se}%
  \BibitemOpen
  \bibfield  {author} {\bibinfo {author} {\bibfnamefont {S.}~\bibnamefont
  {Das}}\ and\ \bibinfo {author} {\bibfnamefont {S.}~\bibnamefont {Ghosh}},\
  }\href {\doibase 10.1103/PhysRevD.80.085009} {\bibfield  {journal} {\bibinfo
  {journal} {Phys. Rev.}\ }\textbf {\bibinfo {volume} {D80}},\ \bibinfo {pages}
  {085009} (\bibinfo {year} {2009})},\ \Eprint {http://arxiv.org/abs/0907.0290}
  {arXiv:0907.0290 [hep-th]} \BibitemShut {NoStop}%
\bibitem [{\citenamefont {Dreiner}\ \emph {et~al.}(2010)\citenamefont
  {Dreiner}, \citenamefont {Haber},\ and\ \citenamefont
  {Martin}}]{Dreiner:2008tw}%
  \BibitemOpen
  \bibfield  {author} {\bibinfo {author} {\bibfnamefont {H.~K.}\ \bibnamefont
  {Dreiner}}, \bibinfo {author} {\bibfnamefont {H.~E.}\ \bibnamefont {Haber}},
  \ and\ \bibinfo {author} {\bibfnamefont {S.~P.}\ \bibnamefont {Martin}},\
  }\href {\doibase 10.1016/j.physrep.2010.05.002} {\bibfield  {journal}
  {\bibinfo  {journal} {Phys.Rept.}\ }\textbf {\bibinfo {volume} {494}},\
  \bibinfo {pages} {1} (\bibinfo {year} {2010})},\ \Eprint
  {http://arxiv.org/abs/0812.1594} {arXiv:0812.1594 [hep-ph]} \BibitemShut
  {NoStop}%
\bibitem [{\citenamefont {Martin}(2010)}]{Martin:1997ns}%
  \BibitemOpen
  \bibfield  {author} {\bibinfo {author} {\bibfnamefont {S.~P.}\ \bibnamefont
  {Martin}},\ }\href {\doibase 10.1142/9789814307505_0001} {\bibfield
  {journal} {\bibinfo  {journal} {Adv.Ser.Direct.High Energy Phys.}\ }\textbf
  {\bibinfo {volume} {21}},\ \bibinfo {pages} {1} (\bibinfo {year} {2010})},\
  \Eprint {http://arxiv.org/abs/hep-ph/9709356} {arXiv:hep-ph/9709356 [hep-ph]}
  \BibitemShut {NoStop}%
\bibitem [{\citenamefont {Britto}(2011)}]{britto1}%
  \BibitemOpen
  \bibfield  {author} {\bibinfo {author} {\bibfnamefont {R.}~\bibnamefont
  {Britto}},\ }\href@noop {} {\enquote {\bibinfo {title} {Qcd and the
  spinor-helicity formalism},}\ } (\bibinfo {year} {2011}),\ \bibinfo {note}
  {introduction to scattering amplitudes Lect1}\BibitemShut {NoStop}%
\bibitem [{\citenamefont {Freidel}\ and\ \citenamefont
  {Hnybida}(2014)}]{Freidel:2013fia}%
  \BibitemOpen
  \bibfield  {author} {\bibinfo {author} {\bibfnamefont {L.}~\bibnamefont
  {Freidel}}\ and\ \bibinfo {author} {\bibfnamefont {J.}~\bibnamefont
  {Hnybida}},\ }\href {\doibase 10.1088/0264-9381/31/1/015019} {\bibfield
  {journal} {\bibinfo  {journal} {Class. Quant. Grav.}\ }\textbf {\bibinfo
  {volume} {31}},\ \bibinfo {pages} {015019} (\bibinfo {year} {2014})},\
  \Eprint {http://arxiv.org/abs/1305.3326} {arXiv:1305.3326 [math-ph]}
  \BibitemShut {NoStop}%
\end{thebibliography}%

\end{document}